\documentstyle[stwol,epsfig]{article}

\begin{document} 
\textheight 20.2cm
\title{\Large\rm
A study of the $\pi^0\pi^0$ system produced  in charge exchange and\protect\\
central collisions}
%
\author{Andrei A. Kondashov$^{\,\rm{a}}$}
\address{\rm $^{\rm{a}}$IHEP, Protvino, Moscow region, 142 284 RUSSIA}
\author{On behalf of the GAMS and WA102 Collaborations}
%
\twocolumn[\maketitle\abstracts{
A study of the $\pi^0\pi^0$ 
system produced in charge exchange $\pi^-p$  collisions at 38 and 
100 GeV/c and in central 
$pp$ interactions at 450~GeV/c
has been carried out. The $S$ wave has rather a complicated structure 
in both processes indicating
the existence of several scalar resonances.
The $f_0(980)$ 
and $f_0(1500)$ appear
as dips at 1 and 1.5 GeV in the $S$ wave for charge exchange reaction,
and as shoulders at these masses in the $S$ wave
for central production. The production of the $f_0(980)$, 
$f_0(1300)$ and $f_0(1500)$ in the reaction $pp \to p_f\pi^0\pi^0p_s$
as a function of the $dP_{T}$ kinematical filter shows the behaviour 
differed from what has been observed for the undisputed $q\bar{q}$ mesons. 
An extra $f_0(2000)$ state
is seen in the $S$ wave for charge exchange reaction as a dip
at 2~GeV. Resonances with higher spins,
$f_2(1270)$, $f_4(2050)$ and $f_6(2510)$, have also been studied.
All the three mesons are produced in the reaction $\pi^-p\to\pi^0\pi^0{n}$
mainly via an one-pion exchange  for small $-t$,
whereas a natural-parity exchange domimates  for large $-t$.
The behaviour of the centrally produced $f_2(1270)$ as a function of the 
$dP_{T}$ is consistent with what has been observed for other 
$q\bar{q}$ states.}]

\section*{1.~INTRODUCTION}

One of the most intriguing problems of the modern meson spectroscopy is
the search for the gluon bound states (glueballs).  A mass of the 
lightest scalar glueball should lie in the region $1500-1750$~MeV 
as expected from the lattice QCD calculations~\cite{bali,sexton}. 
Scalar glueball candidates have been observed in several experiments.
An analysis of the GAMS data on the $IJ^{PC}=00^{++}$
states together with the data of other experiments~\cite{sim} revealed
the existence of five
scalar resonances  in the mass range up to 1.9~GeV. 
One of these states is extra for $q\bar{q}$ systematics,
being a good candidate for the lightest scalar glueball.

Nevertheless, despite on considerable experimental and theoretical efforts
the understanding of the scalar meson sector is rather controversial now.
To identify unambiguously the scalar glueball, the structures of the 
scalar $q\bar{q}$-nonets need to be clarified, 
because
glueball state should be  superfluous for $q\bar{q}$-systematics.
The picture may be complicated if glueball is located in the vicinity 
of $q\bar{q}$-mesons with identical quantum numbers that raises their
mixing and leads to the dispersing of the glueball component over several
mesons.
  
The $\pi^0\pi^0$ system looks very attractive  for experimental 
investigation and for study of the scalar resonances, in particular. 
Only even $J^{PC}$ waves are present in this system, which
simplifies greatly analysis and eliminates  contributions from the
odd waves as compared to the $\pi^+\pi^-$ system. Study of the $\pi^0\pi^0$ 
$S$ wave  in different processes should 
extend  our knowledge about the scalar mesons  
and help to identify  the  states with
enhanced gluonic component. 
\textheight 21cm

In the present report overview of results on the 
$\pi^0\pi^0$ system produced
in charge exchange reaction
\begin{equation}
\begin{array}{c}
\setlength{\unitlength}{1mm}
\begin{picture}(45,9) 
\put(2,7){$\pi^-p \rightarrow M^on$}
\put(15.5,5){\line(0,-1){4}} 
\put(15.5,0){$\rightarrow \pi^o\pi^o \rightarrow 4\gamma$} 
\end{picture}\label{re:cex}
\end{array}
\end{equation}
obtained by the GAMS Collaboration at 38 and 100~GeV/c
is given 
and new GAMS and WA102 
results on the  $\pi^0\pi^0$ system produced in the reaction 
\begin{equation}
\begin{array}{c}
\setlength{\unitlength}{1mm}
\begin{picture}(45,10) 
\put(2,7){$pp \rightarrow p_fM^op_s$}
\put(16,5){\line(0,-1){4}} 
\put(16,0){$\rightarrow \pi^o\pi^o \rightarrow 4\gamma$} 
\end{picture}\label{re:cent}
\end{array}
\end{equation}
at 450~GeV/c are presented. The subscripts $f$ and $s$ in (\ref{re:cent})
indicate the 
fastest and slowest particles in the laboratory frame, respectively.

\section*{2.~DATA SELECTION}

The reaction (\ref{re:cex}) data have been obtained in two 
experiments carried out with the 
GAMS-2000 multiphoton spectrometer in 38~GeV/c $\pi^-$ beam 
extracted from the 70~GeV IHEP proton accelerator (experiment SERP-E-140
at IHEP, Protvino) and  with
the GAMS-4000 spectrometer in 100~GeV/c $\pi^-$ beam of SPS (experiment
NA12 at CERN). The general layout of the experiments,
details of the GAMS-2000 and GAMS-4000 constructions and calibrations
as well as event selection procedures 
have been given in previous publications~\cite{pipi-38-2,pipi-100}. 

\begin{figure}[t]
\center
\epsfig{figure=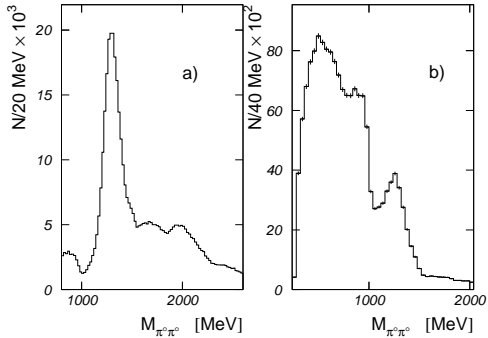,height=1.9in}
\caption{
Invariant mass spectra of the $\pi^0\pi^0$ system
produced in reaction~(\ref{re:cex}) at
100~GeV/c, 
$-t < 0.2$~(GeV/c)$^2$ (a), and in reaction~(\ref{re:cent}), 
WA102 data (b).}
\label{fi:fig-1}
\end{figure}



After kinematical
analysis (3C fit, masses of recoil neutron and two mesons being fixed)
a total of $1.5\times10^6$ and
$6.5\times10^5$ $\pi^0\pi^0$ events are selected at 38 and 100~GeV/c,
respectively.
Mass spectrum of the $\pi^0\pi^0$ events  
for 4-momentum transfer squared $-t < 0.2$~(GeV/c)$^2$
is shown in fig.~\ref{fi:fig-1}a. 
It is dominated by the $f_2(1270)$. A narrow dip 
corresponding to the $f_0(980)$ is clearly seen at 1~GeV. 
A peak  at 2~GeV is identified with
the $f_4(2050)$. A bump around 1.7~GeV is due to the $S$ wave 
contribution (see sect.~4). For  
$-t > 0.2$~(GeV/c)$^2$, the $f_2(1270)$ peak is also clearly seen, 
whereas a shoulder appears on the place of the dip at 1~GeV.

The reaction (\ref{re:cent}) data
come from the  NA12/2 and WA102 experiments
performed in 450~GeV/c proton beam of SPS at CERN. The NA12/2 experiment
has been carried out with the GAMS-4000 spectrometer. 
The WA102 experiment has been performed using CERN 
Omega Spectrometer \cite{omega} and GAMS-4000. 


Separation of the $\pi^0\pi^0$ events is carried out on the basis
of kinematical analysis (6C fit, four-momentum conservation being used and 
masses of two mesons being fixed). Events containing a fast $\Delta^+(1232)$ 
are removed by imposing a cut $M(p_{f}\pi^0)>1.5$~GeV, which leave 55 000
centrally produced $\pi^0\pi^0$ events for NA12/2  and
166 000 $\pi^0\pi^0$ events for WA102.

Mass spectrum of the centrally produced $\pi^0\pi^0$ system 
is shown in fig.~\ref{fi:fig-1}b. A peak at 1.25~GeV corresponding 
to the $f_2(1270)$ and a shoulder at 1~GeV which appears due to the 
interference of the $f_0(980)$ with the $S$ wave background   
are clearly seen. 

 
\section*{3.~PARTIAL WAVE ANALYSIS}

For reaction (1), the coordinate system axes
are defined in the Gottfried-Jackson frame. 
For reaction (2), the $z$-axis is chosen to be 
along the direction 
of the exchanged particle momentum in ``slow'' vertex 
in the $\pi^0\pi^0$ centre of mass system, 
the $y$-axis is defined to be along cross product of 
the exchanged particle momenta in the $pp$ centre of mass system.

The amplitudes used for PWA are defined in the reflectivity basis 
\cite{chung}.
Only amplitudes with spin $z$-projections $|m|=0$ and 
1 are taken into account since amplitudes with $|m|>1$ are equal 
to zero within the error bars as follows from analysis of the angular 
distributions. 
The amplitudes with spin $l=0$, 2 and 4 ($S$, $D$ and $G$ waves, respectively)
are used for the reaction (\ref{re:cex}) PWA at 38~GeV/c, 
the amplitudes with $l=6$ ($J$ waves)
are added at 100~GeV/c. Only $S$ and $D$ waves are considered 
for reaction (\ref{re:cent}).  Contribution of the higher waves is
negligibly small in the mass ranges under study for each reaction.


\begin{figure}[t]
\center
\epsfig{figure=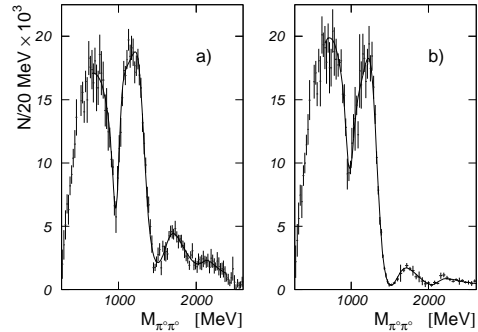,height=1.9in}
\caption{
The $|S|^2$ for 
the $\pi^0\pi^0$ system
produced in reaction~(\ref{re:cex}) at 38 (a) and
100~GeV/c (b), 
$-t < 0.2$~(GeV/c)$^2$. The curves show fit with the
sums of  four relativistic Breit-Wigner functions and 
backgrounds.
}
\label{fi:fig-4}
\end{figure}

\section*{4.~REACTION $\pi^-p\to\pi^0\pi^0n$}

The $S$ wave amplitude module squared
for the physical solution found for $-t < 0.2$~(GeV/c)$^2$
\cite{pipi-100,pipi-38-3}
is shown in fig.~\ref{fi:fig-4}.
The $S$ wave has rather a complicated structure. 
It demonstrates a series of bumps separated with dips at 1,
1.5 and 2~GeV.  The two first dips point to the existence of 
the $f_0(980)$ and $f_0(1500)$ resonances. Rapid variation of the relative 
phase of the $S$ and $D_0$ waves
at 1 and 1.5~GeV confirms the presence of these resonances. 
The dip at 2~GeV is clearly seen at 100~GeV/c,
it is less prominent at 38~GeV/c due to insufficient 
detection efficiency at high mass. This dip  
indicates the presence of a new scalar resonance 
around 2~GeV. Such conclusion is confirmed by the fast variation 
of the $S$ wave phase in this mass region~\cite{pipi-100}. 

A simultaneous $K$-matrix analysis of the GAMS data on the $S$ wave in the  
$\pi^0\pi^0$, $\eta\eta$ and $\eta\eta^{\prime}$ systems produced in charge
exchange reactions at 38~GeV/c in the mass range below 1.9~GeV 
together with the Crystal Barrel, CERN-M\"unich and BNL data
\cite{sim} points to the existence  of four 
comparatively narrow scalar resonances $f_0(980)$, $f_0(1300)$, $f_0(1500)$ 
and $f_0(1780)$ and one broad scalar state $f_0(1530)$
with a width of about 1~GeV. The poles of
the partial amplitude corresponding to physical states
are determined by the mixture of input (``bare'')
states related to the $K$-matrix poles via their transition into real 
mesons. The analysis \cite{sim} shows that 
one bare state in the mass region $1.2-1.6$ GeV 
is superfluous for the $q\bar{q}$-classification, being a good candidate 
for the lightest scalar glueball.  This superfluous bare state is dispersed 
over neighbouring physical states: the narrow $f_0(1300)$ and $f_0(1500)$
resonances and the broad $f_0(1530)$. 
  
\begin{figure}[t]
\center
\epsfig{figure=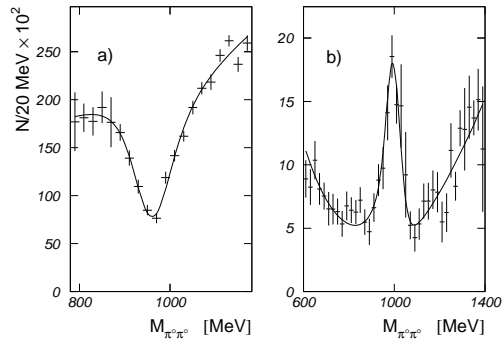,height=1.9in}
\caption{
The $|S|^2$ for 
the $\pi^0\pi^0$ system
produced in reaction~(\ref{re:cex}) at 38~GeV/c, 
$-t < 0.2$~(GeV/c)$^2$ (a) and $-t > 0.3$~(GeV/c)$^2$ (b). 
The curves show fit with the
sums of  relativistic Breit-Wigner functions and polynomial
backgrounds.
}
\label{fi:fig-5}
\end{figure}

For  $-t > 0.3$~(GeV/c)$^2$, a narrow peak 
is seen in the $S$ wave on the place of the dip at 1~GeV observed 
at low momentum transfer (fig.~\ref{fi:fig-5})
\cite{pipi-38-3,pipi-38-1}. 
A mass $997\pm5$ MeV and a width $48\pm10$ MeV of the peak are in good
agreement with the tabulated $f_0(980)$ parameters \cite{PDG}.

A simultaneous analysis of the GAMS data on the $\pi^0\pi^0$ 
$S$ wave around 1~GeV together with the Crystal Barrel and 
CERN-M\"unich data \cite{akssp}
shows that the $f_0(980)$ is strongly related  to the 
$\pi\pi$ channel and much weaker to the
$K\bar{K}$ channel (ratio of $f_0(980)$
coupling constants squared to the $\pi\pi$ and $K\bar{K}$ channels is 
equal to 6). This fact, together with the evidence for the hard component
in the $f_0(980)$ at high $-t$, 
makes unconvincing the interpretation of this 
scalar meson as  a $K\bar{K}$ molecule. 


\begin{figure}[t]
\center
\epsfig{figure=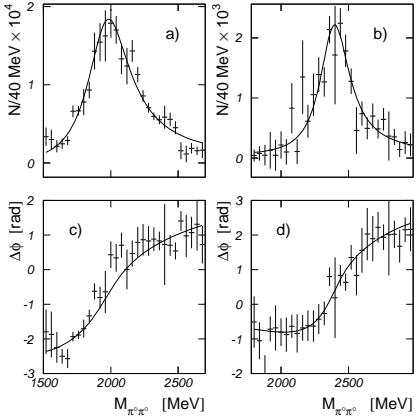,height=2.3in}
\caption{
The $|G_0|^2$ (a) and $|J_0|^2$ (b)  
and phases of the $G_0$ (c) and $J_0$ (d) waves relative
to the $D_0$ wave phase for the $\pi^0\pi^0$ system
produced in reaction~(\ref{re:cex}) at 100~GeV/c, 
$-t < 0.2$~(GeV/c)$^2$. The curves show fit with the
sums of relativistic Breit-Wigner functions and polynomial
backgrounds. 
}
\label{fi:fig-7}
\end{figure}


Mesons with higher spins, $f_2(1270)$, $f_4(2050)$ and $f_6(2510)$,
are seen as clear peaks in the $D$, $G$ and $J_0$ waves, respectively
(fig.~\ref{fi:fig-7}). For $-t < 0.2$~(GeV/c)$^2$, all the three 
mesons are produced
via an one pion exchange with a small absorption.
Ratios of the $f_2(1270)$ amount in the 
$D$ waves with $|m| = 0$ and 1  are equal to 
$7\%$ at 38~GeV/c and $3\%$ at 100~GeV/c \cite{pipi-100,pipi-38-1}. 
The ratios of the $f_4(2050)$ amount in the  $G_0$ and $G_{\pm}$ waves 
are the same as those  for the $f_2(1270)$ in the $D$ waves. 
As for the $f_6(2510)$, the upper limit is set for its production in
the $J_{\pm}$ waves as compared to the $J_0$ wave ($<0.1$, $95\%$ C.L.)
\cite{pipi-100}.

%


With momentum transfer increase, an un\-na\-tu\-ral\--pa\-ri\-ty 
exchange is died away.
For $-t > 0.3$~(GeV/c)$^2$, the $f_2(1270)$ and $f_4(2050)$ 
are produced predominantly 
via a natural-parity
exchange ($D_+$ and $G_+$ waves). This shows a similarity of the production
mechanisms of the $f_2(1270)$ and $f_4(2050)$.


\begin{figure}[t]
\center
\epsfig{figure=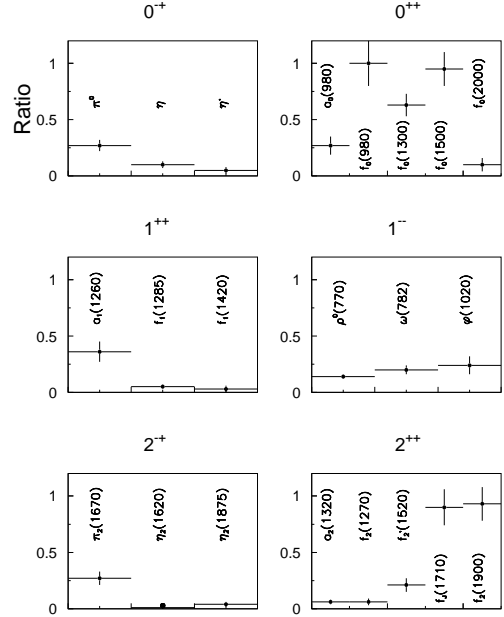,height=3.4in}
\caption{
The ratios of the amounts of resonances with $dP_T < 0.2$~GeV/c to the
amounts with $dP_T > 0.5$~GeV/c.
}
\label{fi:fig-00}
\end{figure}

\section*{5.~A KINEMATICAL $dP_T$ FILTER}

Production of the states with valence gluons
may be enhanced using glue-rich production mechanisms. One such mechanism
is Double Pomeron Exchange (DPE) where the pomeron is thought to be a
multi-gluonic object. With increasing energy, 
Double Pomeron Exchange (DPE) becomes
relatively more important in central production 
with respect to other
exchange processes (Reggeon-Pomeron and Reggeon-Reggeon exchange) 
\cite{kirk}. 
 
Recently it has been proposed~\cite{filter,close} 
to analyse the centrally produced resonances in terms of the 
difference in transverse momenta of the exchanged particles which
is defined as follows
\begin{equation}
dP_T = \sqrt{(P_{y_1}-P_{y_2})^2+(P_{z_1}-P_{z_2})^2}
\end{equation}
where $P_{y_i}$, $P_{z_i}$ are the $y$ and $z$ components of the momentum
of the $i$-th exchanged particle in the $pp$ centre of mass system.
It has been observed that all the undisputed $q\bar{q}$ states 
(i.e. $\rho^0(770)$, $\eta^{\prime}$, $f_1(1285)$, $f_2(1270)$,
$f_2^{\prime}(1525)$ etc.) are suppressed at small $dP_T$, whereas the 
glueball candidates $f_0(1500)$, $f_J(1710)$ and $f_2(1900)$ survive. 
 
Figure \ref{fi:fig-00} shows the ratios of the event numbers for 
different resonances for small and large $dP_T$  found from the fit
to the efficiency corrected mass spectra. It is clearly seen that all
the undisputed $q\bar{q}$ states which can be produced in DPE
have very small values for this ratio  ($\le0.1$).
Some states which can not be produced by DPE (namely those with negative
$G$ parity or $I=1$) have slightly higher values ($\approx0.25$). However, all
of these states are suppressed relative to the non-$q\bar{q}$ candidates
$f_0(980)$, $f_0(1300)$, $f_0(1500)$, $f_J(1710)$ and $f_2(1900)$ 
which have values for this ratio of about~1.

\section*{6.~REACTION $pp\to{p_f}\pi^0\pi^0p_s$}

A PWA of the reaction $pp\to{p_f}\pi^0\pi^0p_s$ has been carried
out in the mass range from the threshold up to 1.8~GeV. 
The $S$ wave amplitude for one of the two PWA solutions 
is much smaller than amplitudes of the $D$ waves in the whole
mass range. This solution is rejected as unphysical one.

\begin{figure}[t]
\center
\epsfig{figure=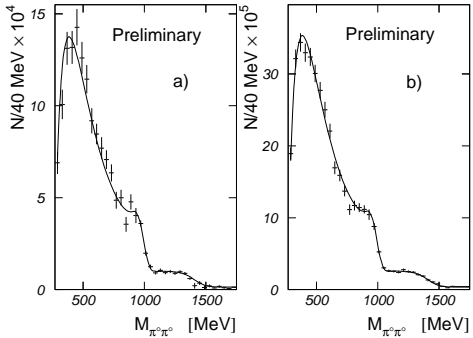,height=1.9in}
\caption{
The $|S|^2$ for 
the $\pi^0\pi^0$ system
produced in reaction~(\ref{re:cent})  in the 
NA12/2 (a) and WA102 (b) experiments.
The curves show fit with the  
sums of  two relativistic Breit-Wigner functions and 
backgrounds.
}
\label{fi:fig-8}
\end{figure}

The $S$ wave for the physical solution is characterized by a broad bump 
below 1~GeV and two shoulder, at 1 and 1.4~GeV (fig.~\ref{fi:fig-8}). 
The first shoulder is attributed to the $f_0(980)$, the second one
may be explained by the interference of the $f_0(1300)$ and $f_0(1500)$
resonances (see sect.~6). Peaks 
corresponding to the $f_2(1270)$ are seen in the $D_0$ and $D_-$ waves,
such peak is absent in the $D_+$ wave. 
Ratio of the $D_-$ and $D_0$ wave intensities is about $20\%$
at the $f_2(1270)$ mass. 


In order to apply the kinematical $dP_T$ filter,
PWA have been performed in intervals $dP_T < 0.35$~GeV/c,
$0.35 < dP_T < 0.6$~GeV/c and $dP_T > 0.6$~GeV/c.
The shoulders at 1 and 1.4~GeV in the $S$ wave 
have approximately the same heights in the three  $dP_T$ intervals, 
whereas the bump below
1~GeV becomes much more prominent for small 
$dP_T$. The $f_2(1270)$ peaks are clearly seen
in the $D_0$ and $D_-$ waves for large $dP_T$ 
and disappear for small $dP_T$.

\vskip 2mm
\footnotesize
\noindent
Table 1. \\
Resonance production as a function of $dP_T$ expressed 
as a percentage of its total contribution. $dP_T$ intervals are
given in~GeV/c.\\[-10mm]

\noindent
\begin{center}
\begin{tabular}{cccc}
\hline\\
&$dP_T < 0.35$&$0.35 < dP_T < 0.6$&$dP_T > 0.6$ \\
\hline
\\
$f_0(980)$  & $34\pm7$ & $42\pm7$ & $24\pm5$ \\
$f_0(1300)$ & $30\pm9$ & $38\pm8$ & $32\pm7$ \\
$f_0(1500)$ & $32\pm8$ & $42\pm7$ & $26\pm7$ \\
$f_2(1270)$ & not seen & $24\pm5$ & $76\pm4$ \\
\hline
\end{tabular}
\end{center}
\normalsize
\vskip 2mm

Relative contributions of the resonances observed in the centrally produced 
$\pi^0\pi^0$ system for three $dP_T$ intervals have been estimated from a
simultaneous fit to the NA12/2 and WA102 data (see sect.~7). 
These contributions are presented in table 1.
The $f_0(1300)$ and $f_0(1500)$ have a similar behaviour as a function of
the $dP_T$  not consistent with that observed for $q\bar{q}$ states. 
It is interesting to note that the enigmatic $f_0(980)$ also 
does not behave as a normal $q\bar{q}$ state. In contract, the $f_2(1270)$ 
is suppressed for small $dP_T$ and enhanced for large $dP_T$ in agreement with 
the behaviour of the other $q\bar{q}$ states.

\section*{7.~FIT TO THE $S$ WAVE}

In order to determine the parameters of the scalar resonances, 
a fit to the $S$ wave has been performed.
The following parametrisation is used:
\begin{equation}
A(M_{\pi\pi})  =  G(M_{\pi\pi}) +
           \sum_{n=1}^{N_{res}} a_n e^{i\theta_n} B_n(M_{\pi\pi}),  
\label{eq:samp}
\end{equation}
\begin{equation}
G(M_{\pi\pi})  =  (M_{\pi\pi}-2m_{\pi})^{\alpha}
e^{-\beta{M_{\pi\pi}}-\gamma{M^2_{\pi\pi}}}
\end{equation}
where $a_n$ and $\theta_n$ are the amplitude and the 
phase of the $n$-th resonance,
respectively, $\alpha$, $\beta$ and $\gamma$ are the real parameters,
$B(M_{\pi\pi})$ is the relativistic Breit-Wigner function.
To describe the $|S|^2$, function (\ref{eq:samp}) module squared has been 
convoluted with a Gaussian to account the experimental 
mass resolution.

First, a fit to the $S$ wave for the $\pi^0\pi^0$ system produced in 
reaction (1) at 38 and 100~GeV/c has been carried out 
using four resonances (fig.~\ref{fi:fig-4}). 
Three of them, $f_0(980)$, $f_0(1500)$ and 
$f_0(2000)$, correspond to the dips at 1, 1.5 and 2~GeV. One more resonance,
$f_0(1300)$, is needed to describe the bump around 1.2~GeV. Without
this resonance, quality of the fit deteriorates significantly.
As a result of the  $f_0(1300)$ and $f_0(1500)$ 
interference  with the $S$ wave background, the mass of the $f_0(1300)$
turns out to be shifted to higher values as compared to the bump maximum,
the $f_0(1500)$ mass is also shifted as compared to the dip position.
Parameters of the
scalar resonances determined from the fit are presented in table~2.

Second, a fit to the  
reaction (2) $S$ wave has been performed 
(fig.~\ref{fi:fig-8}). 
To begin, two resonances are included in the fit to describe the
shoulders at 1 and 1.4~GeV. A mass $988\pm10$~MeV and 
a width $76\pm20$~MeV of the first resonance are consistent
with the tabulated parameters of the $f_0(980)$ \cite{PDG}.
For the second resonance a mass of $1420\pm30$~MeV and 
a width of $230\pm50$~MeV are obtained. This state can be 
reproduced as a result  of the  $f_0(1300)$ and $f_0(1500)$ interference.
The amount of the $f_0(1300)$ is found to be about $20\%$ of that
of the $f_0(1500)$. 
  
\textheight 20cm

Finally, a simultaneous fit to the 
reaction (1) and (2) $S$ wave has been performed. 
The $f_0(980)$, $f_0(1300)$, $f_0(1500)$ and $f_0(2000)$
are introduced to describe the $S$ wave in reaction (1), only first three
resonances are used in the fit to the $S$ wave in reaction (2).
Fit gives the  $f_0(980)$ and $f_0(1300)$ masses
consistent with those obtained from the fit to the charge exchange data,
whereas a mass of the $f_0(1500)$ is shifted 
to lower values. The $f_0(1500)$ mass
found from the simultaneous fit
agrees well with the tabulated value \cite{PDG}, while the width
turns out to be somewhat larger. As a result of the $f_0(1500)$ mass shift,
a mass of the $f_0(2000)$
also becomes slightly smaller but agrees within 
the errors with the value obtained from the reaction (1) $S$ wave fit.
The state with a similar  mass and width
was observed by the WA102 Collaboration  in the reaction $pp\to
p_f(\pi^+\pi^-\pi^+\pi^-)p_s$~\cite{om-4pi}. 
   
\vskip 2mm
\footnotesize
\noindent
Table 2.\\
Parameters (in MeV) 
of the scalar resonances obtained from the fit
to the reaction (1) $S$ wave 
(fit I) and  the simultaneous fit to the
reactions (1) and (2) $S$ wave (fit~II).\\[-5mm]

\noindent
\begin{center}
\begin{tabular}{ccccc}
\hline\\
&
\multicolumn{2}{l}{~~~Fit I}&
\multicolumn{2}{l}{~~~Fit II} \\
\cline{2-5}\\
 & Mass & Width &  Mass & Width  \\
\hline
\\
$f_0(980)$  & $970\pm10$  & $85\pm20$ & $980\pm10$ & $80\pm20$ \\
$f_0(1300)$ & $1310\pm25$ & $195\pm40$ & $1300\pm25$ & $220\pm40$ \\ 
$f_0(1500)$ & $1590\pm80$ & $300\pm90$ & $1495\pm35$ & $250\pm60$ \\
$f_0(2000)$ & $2020\pm60$ & $220\pm80$ & $1960\pm60$ & $210\pm80$ \\
\hline 
\end{tabular}
\end{center}
\normalsize

\section*{8.~CONCLUSIONS}

The partial wave analyses of the $\pi^0\pi^0$ system produced in
charge exchange $\pi^-p$ reaction at 38 and 100~GeV/c and in central 
$pp$ collisions at 450~GeV/c have been carried out.
The $f_0(980)$ and $f_0(1500)$ appear in different ways in the two
processes. These resonances 
are observed as  dips in the $S$ wave at 1 and 1.5~GeV in 
charge exchange reaction, whereas 
in central production the $f_0(980)$ and $f_0(1500)$ are seen as 
shoulders. The $f_0(1300)$ is essential for the description
of charge exchange data while in central production contribution 
of this resonance is less prominent. The centrally produced  
$f_0(980)$, $f_0(1300)$ and  $f_0(1500)$ have a similar dependence 
as a function of the $dP_T$, differed from that observed 
for all the undisputed $q\bar{q}$ mesons.  

An extra $f_0(2000)$ state is observed in charge exchange reaction. 
It is similar to the scalar resonance observed by the WA102 Collaboration 
in the reaction $pp\to p_f(\pi^+\pi^-\pi^+\pi^-)p_s$. 
Existence  of a new scalar state around 2 GeV
may be very 
important for understanding of the scalar nonet structure and 
for isolating the lightest scalar glueball. 

For mesons with higher spins, production mechanisms of the $f_2(1270)$,
$f_4(2050)$ and $f_6(2510)$ as a function of momentum transfer 
in reaction (1) have been studied. All the three mesons are produced via
a dominating one-pion exchange for small $-t$, whereas for large
$-t$ the $f_2(1270)$ and $f_4(2050)$ are produced predominantly
through a natural-parity $t$-channel exchange.
The $f_2(1270)$ is clearly seen in the reaction (2) 
$D_0$ and $D_-$ waves for large $dP_T$.
Its behaviour as a function of the $dP_T$ is consistent with that expected 
for $q\bar{q}$ state.      

%

\end{document}